# Non-reciprocal magnetoresistance, directional inhomogeneity and mixed symmetry Hall devices.

G. Kopnov and A. Gerber*

Raymond and Beverly Sackler Faculty of Exact Sciences,

School of Physics and Astronomy, Tel Aviv University,

Ramat Aviv, 69978 Tel Aviv, Israel

Phenomenology similar to the non-reciprocal charge transport violating Onsager's reciprocity relations can develop in directionally inhomogeneous conducting films with non-uniform Hall coefficient along the current trajectory. The effect is demonstrated in ferromagnetic CoPd films and analyzed in comparison with the unidirectional magnetoresistance phenomena. We suggest to use an engineered inhomogeneity for spintronics applications and present the concept of mixed symmetry Hall devices in which transverse to current Hall signal is measured in a longitudinal contacts arrangement. Magnetization reversal and memory detection is demonstrated in the three-terminal and the partitioned normal metal - ferromagnet (NM-FM) device designs. Multi-bit memory is realized in the partitioned FM-NM-FM structure. The relative amplitude of the antisymmetric signal in the engineered ferromagnetic devices is few percent which is $10 - 10^3$ times higher than in their unidirectional magnetoresistance analogues.

* Corresponding author. Email: gerber@tauex.tau.ac.il

Non-reciprocal charge transport violating Onsager's reciprocity relations [1] attracts growing interest in recent years [2, 3]. Observation of longitudinal resistance composed of even $R_{xx,even}$ and odd $R_{xx,odd}$ in magnetic induction contributions, such that $R_{xx}(B) \neq R_{xx}(-B)$, has been reported in a variety of non-centrosymmetric quantum materials [4], non-centrosymmetric superconductors [5], ferromagnet/ normal metal bilayers [6 - 8], magnetic topological insulators [9, 10], topological insulator/ferromagnet heterostructures [11] and semiconducting crystals [12 - 14]. The so-called unidirectional spin Hall magnetoresistance (USMR) discovered in Pt/Co and Ta/Co bilayers [6] is a nonlinear and nonreciprocal effect that modulates the longitudinal resistivity depending on the component of the in-plane magnetization vector. It was understood as resulting from the interaction of the current-induced interface spin accumulation due to the spin Hall effect in a heavy normal metal layer and magnetization of the ferromagnetic layer. The resistance is minimum if the spin at the interface and the magnetization are parallel, whereas it is maximum if they are antiparallel. From the application point of view, the effect allows to differentiate between opposite magnetization orientations in simple two-terminal FM/NM bi-layer structures with no need in an additional fixed magnetization layer required for the spin-valve spintronic devices [15]. It was further suggested to use the USMR in FM/NM/FM trilayers to construct two-bit memory devices [16]. Non-trivial in its physical origins, the USMR has practical limitations being relatively small in magnitude $R_{xx,odd}/R_{xx,even} \sim 10^{-5}$ and requiring high current density for operation ( > $10^7$ A/cm$^2$) [6]. Electron-magnon scattering contribution of the effect can be larger than ~ 5×10$^{-4}$ at low field, increasing quadratically with current density [17, 18]. Search for a better performance continues [8] and higher efficiency up to 0.2% was reported in a paramagnet-(Ga,Mn)As/ferromagnet-(Ga,Mn)As bilayer [7].

A different mechanism with a similar antisymmetric phenomenology can develop in directionally inhomogeneous conducting materials with Hall coefficient varying along the current trajectory. Following Segal et al. [19] a sample in a four-probe geometry, such as Hall bar, can be represented by a simple circuit shown in Fig. 1. $V_{xy,ab}$ and $V_{xy,cd}$ indicate the transverse voltage generated by the Hall effects at two cross-sections ab and cd respectively, while two equal resistors $2R$ are positioned between points a and c, and b and d (the total resistance is $R$). According to Kirchhoff's circuit laws, the longitudinal voltages along the left and right edges of the bar are:

$$V_{ac} = IR + \frac{(V_{xy,ab} - V_{xy,cd})}{2}$$

$$V_{bd} = IR - \frac{(V_{xy,ab} - V_{xy,cd})}{2} \tag{1}$$

Voltage measured at an edge along the current flow differs from the ordinary Ohmic one $IR$ when $V_{xy,ab} \neq V_{xy,cd}$. The field symmetry of $(V_{xy,ab} - V_{xy,cd})$ is even in the case of e.g. non-uniform planar Hall effect contributions and odd when the ordinary and / or extraordinary Hall effects are present. In the latter case, the odd component $V_{xx,odd}$ is given by:

$$V_{xx,odd} = \frac{V_{xy,ab} - V_{xy,cd}}{2} = \frac{1}{2}\mu_0 I \left( \frac{R_{EHE,ab} M_{ab}(H_{ab})}{t_{ab}} - \frac{R_{EHE,cd} M_{cd}(H_{cd})}{t_{cd}} \right) \tag{2}$$

where $t_{ab}$ and $t_{cd}$ are the local thickness, $M_{ab}$ and $M_{cd}$ are the local magnetization depending on the local field $H_{ab}$ and $H_{cd}$, and $R_{EHE,ab}$ and $R_{EHE,cd}$ are the EHE coefficients at cross-sections ab and cd respectively. Variation of any of these parameters along the film due geometrical, material or field inhomogeneity can be the origin of the odd longitudinal term. A particular case in which two domains with antiparallel magnetization and the separating domain wall positioned between the voltage probes has been considered in Ref. [20]. However, as seen in Eq.2, conditions for the development of an odd longitudinal signal are much more general. Predictions of this simple circuit model with only two current channels were confirmed by a more rigorous derivation of the electric potential along an infinitely long sample with variable Hall coefficient and by numerical calculations for a finite sample [19]. The directional inhomogeneity effect doesn't violate the Onsager's reciprocity, therefore is fundamentally different from the true non-reciprocal magnetoresistance. Also, being a superposition of the resistivity and the Hall effect it should not be referred as magnetoresistance. Phenomenologically, the effect can be identified and distinguished from the non-reciprocal magnetoresistance by: (i) $V_{xx,odd} \approx (V_{xy,ab} - V_{xy,cd})/2$ (approximate equality is due to simplifications of the model); (ii) the voltage signal is proportional to current amplitude and polarity, while the USMR is quadratic in current, and (iii) the signal polarity is reversed when measured along an opposite film edge.

Fig. 2 illustrates the directional inhomogeneity effect. The sample is 10 nm thick and 100 µm wide $Co_{20}Pd_{80}$ ferromagnetic film grown by RF sputtering in a Hall bar geometry. The sample has an out-of-plane anisotropy and displays a square hysteresis at 77K. Fig.2a presents the longitudinal voltages $V_{xx,ac}$ and $V_{xx,bd}$ measured parallel to current at two opposite edges between contacts a - c and b - d respectively as a function of field normal to the film plane. Current in this measurement and in the following experiments was 100 µA dc. Antisymmetric in field voltage spikes develop when magnetization reverses. Their amplitude grows linearly with current and polarity reverses when either the current direction is reversed or when the signal is measured along an opposite bar edge. The background signals at two edges differ slightly from each other (left and right vertical axes) which might be the result of either misalignment of the probes or a transverse material inhomogeneity. Fig.2b displays the transverse (EHE) Hall voltage $V_{xy,ab}$ and $V_{xy,cd}$ measured at two cross-sections $ab$ and $cd$ respectively. The signals are not identical: coercive field at $ab$ is slightly higher than at $cd$, thus the Hall voltages differ significantly during reversal of the magnetization (see the inset). Finally, Fig.2c compares the odd component of the longitudinal voltage $V_{xx,odd}$ and the differential transverse signal $(V_{xy,ab} - V_{xy,cd})/2$. The signals match each other, therefore we attribute the effect to the directional inhomogeneity.

Similar antisymmetric magnetoresistance spikes during the magnetization reversal were observed in a variety of materials, such as common metallic and semiconducting ferromagnets [19-22], magnetic topological insulators [23, 24] and non-collinear antiferromagnets [25, 26].

In our case, non-uniformity of the magnetization reversal illustrated in Fig.2 is an unintended artifact of fabrication. On the other hand, similar to the application opportunities of the USMR effect, one can adapt an engineered directional inhomogeneity to develop alternative designs of the Hall effect spintronics devices [27]. Here, we present the concept of the mixed symmetry Hall devices in which transverse to current Hall signals are measured in a longitudinal contacts arrangement. Efficiency of such devices is determined by the ratio between the antisymmetric and symmetric signals: $E = V_{xx,odd}/V_{xx,even}$. The differential transverse voltage $V_{xx,odd} = (V_{xy,ab} - V_{xy,cd})/2$ is the largest when Hall voltage at one cross-section is zero, e.g. when $V_{xy,cd} = 0$. This can be achieved using the (i) three terminal Hall bar or (ii) partitioned FM-NM Hall bar designs. Sketch of a three-terminal device is shown in the inset of Fig.3. The bar has 3 contacts: $a$ and $c$ are used for current injection, and *a* and *b* for voltage measurements (terminal *a* is common). Hall voltage at point *a* is zero, therefore $V_{xx} = V_{xx,ab} + V_{xy.Bb}/2$, where $V_{xx,ab} = IR_{xx}$ is the resistance signal and $V_{xy.bB}$ is the Hall voltage at the cross-section $Bb$. Demonstration of the measured response in the three-terminal device using a 20 nm thick $Co_{20}Pd_{80}$ film is shown in Fig.3. $V_{xx}$ measured at room temperature and at 77K is shown as a function of normal to plane field. The sample displays no hysteresis at room temperature, therefore the signal can be used to detect magnetization reversal and/or to serve as a field sensor in the magnetization reversal range. Negative magnetoresistance beyond the reversal range is due to the spin-magnon scattering [28]. Efficiency of the magnetization reversal detection is: $\frac{(V_{xx,\uparrow} - V_{xx,\downarrow})}{V_{xx,even}} = 0.15\%$ in this case, where $V_{xx,\uparrow}$ and $V_{xx,\downarrow}$ correspond to the up- and down- magnetized states. The sample exhibits a square hysteresis at 77K, which allows the memory read-up. The efficiency in this case is 0.4%. Efficiency of a three terminal device is limited by the Hall angle of the material and geometry of the device. The Hall angle is given by the ratio $\rho_{xy}/\rho_{xx}$, where $\rho_{xy}$ and $\rho_{xx}$ are the EHE and the longitudinal resistivity respectively. This ratio is about 0.8% in 20 nm thick $Co_{20}Pd_{80}$ and 1% in thin films of ordinary ferromagnets [29]. The device efficiency can be increased to few percent by optimizing the length/width aspect ratio and position and size of contacts [30, 31].

An alternative structure of the mixed symmetry device is a partitioned FM-NM Hall bar design shown in inset of Fig.4. The bar is divided into two sections: a ferromagnetic metal (FM) and a normal (NM) one. The NM section is a thick low resistance normal metal film with an ordinary Hall effect signal negligible compared with the magnitude of extraordinary Hall effect in a thin ferromagnetic cross-bar. The device can be operated either in a three- or four-terminal configuration. Fig.4 presents two implementations of such device composed of thin $Co_{20}Pd_{80}$ (FM) and thick Cu/Au (NM) sections. The device was deposited through a 0.5 mm wide stainless steel shadow mask in two steps exposing only part of the substrate at each stage. 20 nm thick $Co_{20}Pd_{80}$ segment was deposited first followed by 100 nm thick Cu/Au section. No field dependent hysteresis is observed at room temperature and, as in a previous case, the signal can be used to detect magnetization reversal and/or to serve as a field sensor in the magnetization reversal range. Resistance of the device is reduced by the factor 5.6 compared with the single FM design (Fig.3) and the efficiency is increased to 0.8%. Open circles present the same sample measured at 77K where the film exhibits hysteresis. Two zero field memory states correspond to the up and down magnetization orientations. Efficiency of this device is 2.1%.

The concept of the partitioned Hall device can be developed further to construct a multi-bit magnetic memory cell. Sketch of the memory unit and its experimental implementation is shown in Fig.5. The device is composed of 3 sections: 2 ferromagnetic cross-bars FM1 and FM2 possessing different perpendicular anisotropies and EHE magnitudes interconnected by a low resistance non-magnetic segment NM. In the present demonstration the ferromagnetic sections are 10 nm and 20 nm thick $Co_{20}Pd_{80}$ films interconnected by 100 nm thick CuAu section deposited in three subsequent steps. EHE resistance of bars FM1 and FM2 as a function of field are shown in Fig.5a. Coercivity of the 20 nm thick bar is higher than that of the 10 nm thick one while the saturated EHE resistance of the latter is the highest. Fig.5b shows the longitudinal voltage obtained as a function of applied magnetic field following the major and minor hysteresis loops. Dashed arrows indicate the field sweep directions. Four zero field voltage states, making a two-bit memory, correspond to 4 combinations of magnetization in two ferromagnetic sections (up-up; down-down, up-down and down-up). Efficiency of the device is 1.25%, defined as $\frac{(V_{xx,\uparrow\downarrow}-V_{xx,\downarrow\uparrow})}{V_{xx,even}}$. Higher order memory, e.g. 4-bit with 16 memory states can be fabricated using two double layer ferromagnetic cross-bars where all layers have distinct coercivities and EHE resistances. Such single unit multi-bit device differs from the split cell MRAM suggested in Ref. [32].

To summarize, voltage developing along an edge of a current carrying film with a directionally non-uniform Hall coefficient contains both an even in magnetic induction component due to the longitudinal resistivity and an odd symmetry term due to a difference in Hall voltages at the probes cross-sections. This effect of the directional inhomogeneity is general and applies to any conducting material, including the ferromagnets discussed here, but also to normal metals, semiconductors and superconductors in their transition range. We suggest to use an engineered inhomogeneity for spintronics applications and present the concept of mixed symmetry Hall devices in which transverse to current Hall signal is measured in a longitudinal contacts arrangement. Feasibility of the magnetization reversal and memory detection using the three-terminal and the partitioned ferromagnet/normal metal bar designs was demonstrated. Multi-bit memory device was realized in the partitioned FM-NM-FM structure. The relative amplitude of the asymmetric signal in the optimized devices is few percent which is $10 - 10^3$ times higher than in their unidirectional magnetoresistance analogues. Possibilities to incorporate the mixed symmetry Hall devices in their lateral or vertical designs within spintronics architecture are open for further studies.

The work was supported by the Israel Science Foundation grant No. 992/17.

The data that supports the findings of this study are available within the article.

**Figure captions.**

Fig.1. Electronic circuit presentation of a current carrying Hall bar. $V_{xy,ab}$ and $V_{xy,cd}$ indicate the Hall voltage at cross-sections ab and cd respectively. Resistance of the circuit is $R$.

Fig.2. Longitudinal and transverse voltages in 10 nm thick and 100 µm wide $Co_{20}Pd_{80}$ Hall bar as a function of normal to plane field at 77 K. $I = 100$ µA dc. (a) - Longitudinal voltage $V_{xx,ac}$ (open circles, left vertical axis) and $V_{xx,bd}$ (solid circles, right vertical axis) measured parallel to current at two opposite edges between contacts a - c and b - d respectively. The contacts labeling is the same as in Fig.1. (b) - Transverse (EHE) Hall voltage $V_{xy,ab}$ and $V_{xy,cd}$ measured at cross-sections $ab$ and $cd$ respectively. Inset: zoom of the magnetization reversal range. (c) – comparison between the odd component of the longitudinal voltage $V_{xx,odd}$ (solid circles, left vertical axis) and the differential transverse signal $\Delta V_{xy}/2 = (V_{xy,ab} - V_{xy,cd})/2$ (open circles, right vertical axis).

Fig.3. Sketch and demonstration of a three-terminal Hall device. The bar has 3 contacts: *a* and *c* are used for current injection, and *a* and *b* for voltage measurements (terminal *a* is common). Solid (red) circles indicate the field dependence of the longitudinal voltage at room temperature and open ones at 77K. The film is 20 nm thick and 100 µm wide $Co_{20}Pd_{80}$.

Fig.4. Sketch and demonstration of a partitioned FM-NM Hall bar device. FM section is 20 nm thick $Co_{20}Pd_{80}$ and NM is 100 nm thick CuAu. Solid (red) circles indicate the room temperature longitudinal voltage as a function of field and open circles are the 77K measurement.

Fig.5. Sketch and demonstration of a partitioned two-bit magnetic memory cell. The device is composed of 3 sections: 2 ferromagnetic cross-bars FM1 (10 nm thick $Co_{20}Pd_{80}$) and FM2 (20 nm thick $Co_{20}Pd_{80}$) interconnected by a low resistance non-magnetic segment NM (100 nm thick

CuAu). (a) – Transverse voltage in bars FM1 and FM2 as a function of field. (b) - Longitudinal voltage as a function of field following the major and minor hysteresis loops. Dashed arrows indicate the field sweep directions. Four zero field voltage states, making a two-bit memory, correspond to 4 combinations of magnetization in two ferromagnetic sections (up-up; down-down, up-down and down-up). T = 77K.

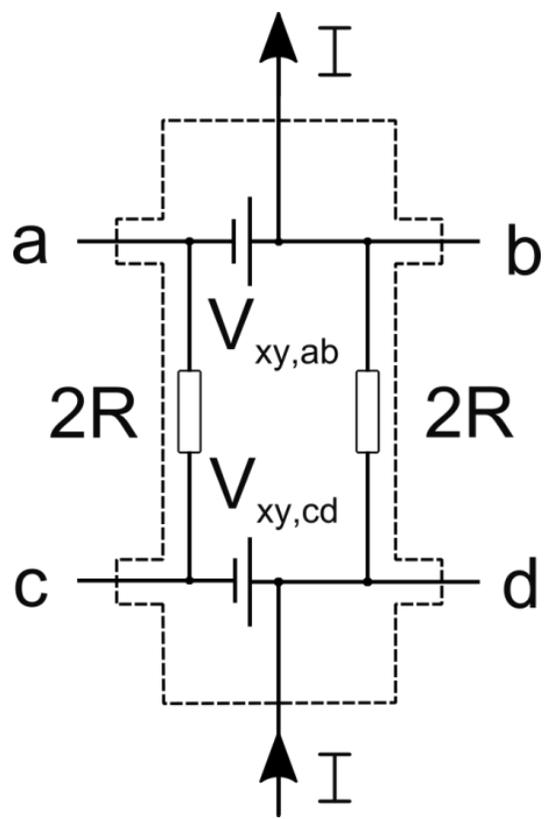

Fig.1

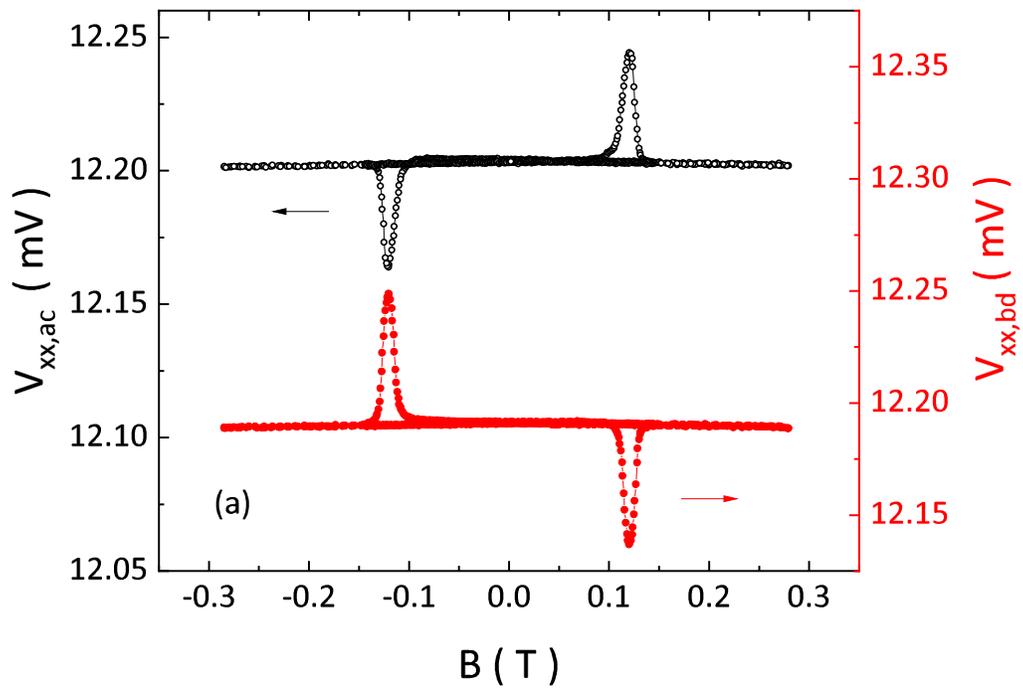

Fig. 2a

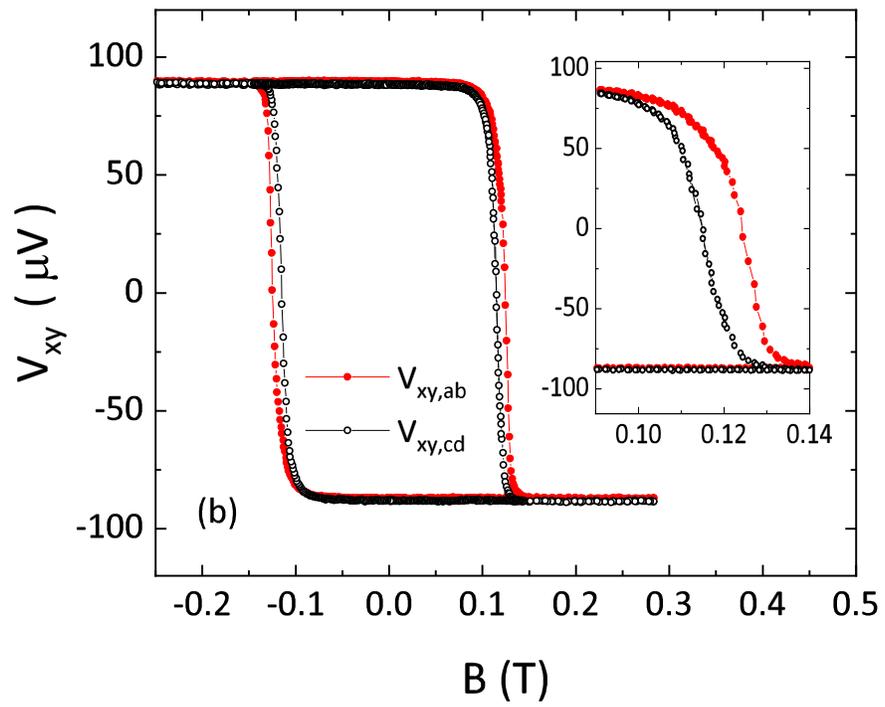

Fig. 2b

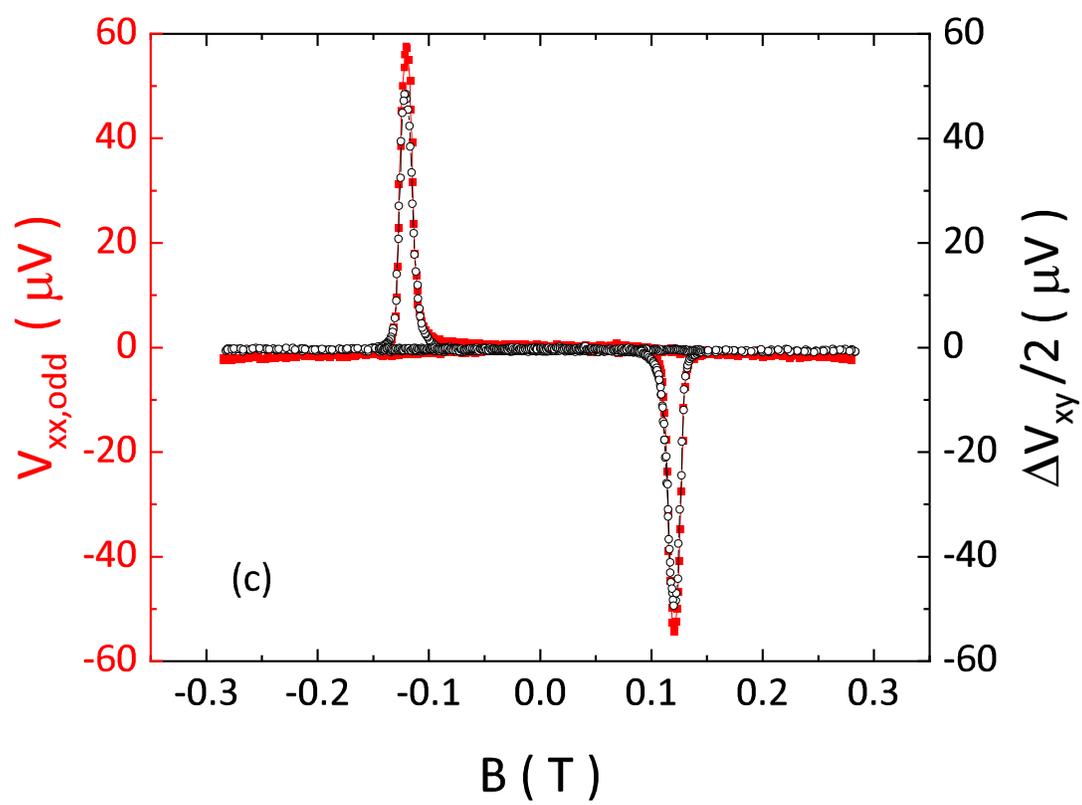

Fig.2c

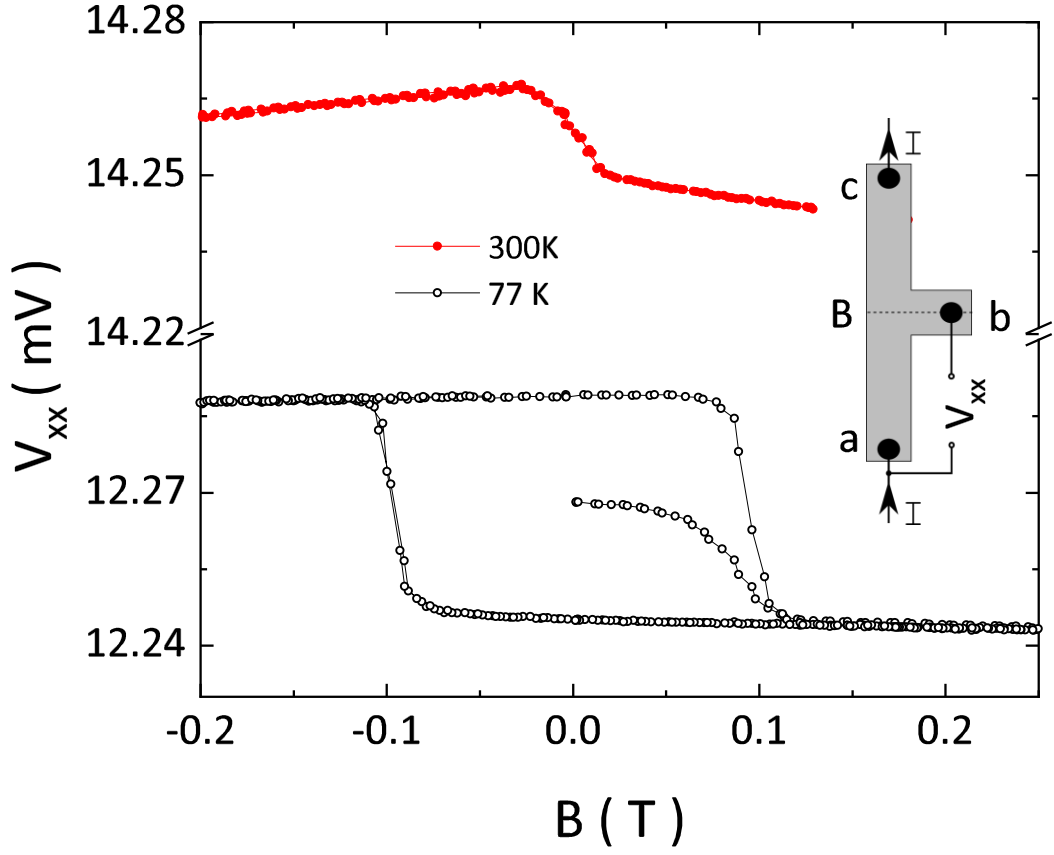

Fig.3

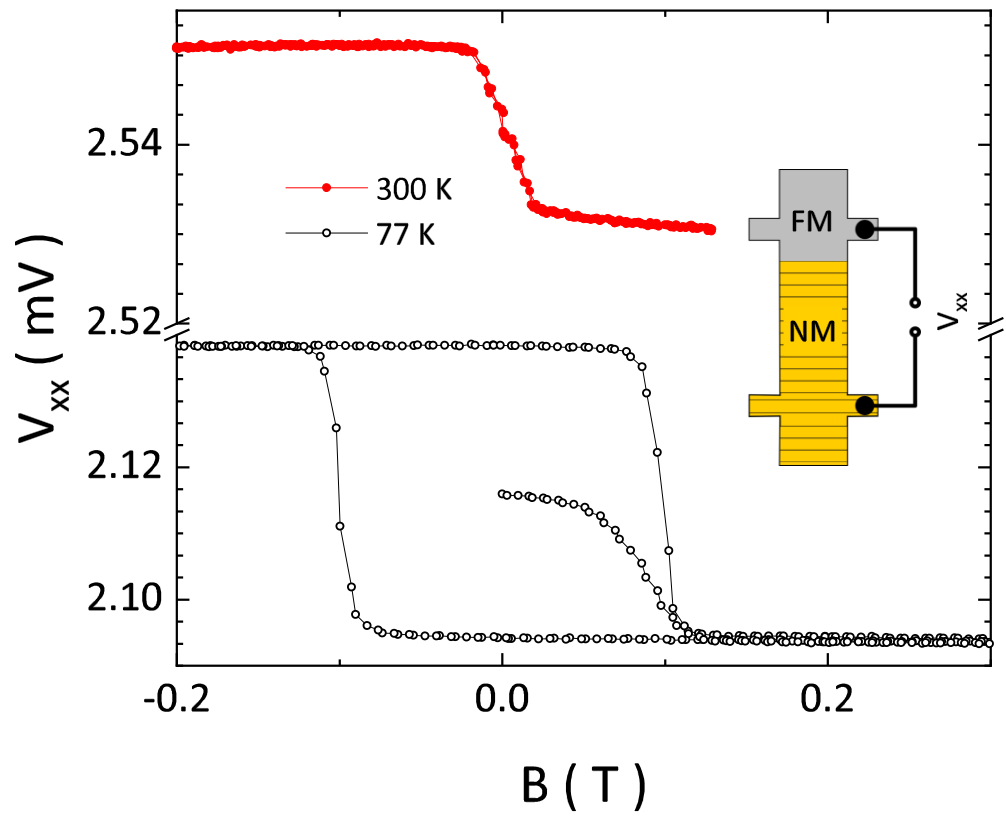

Fig.4

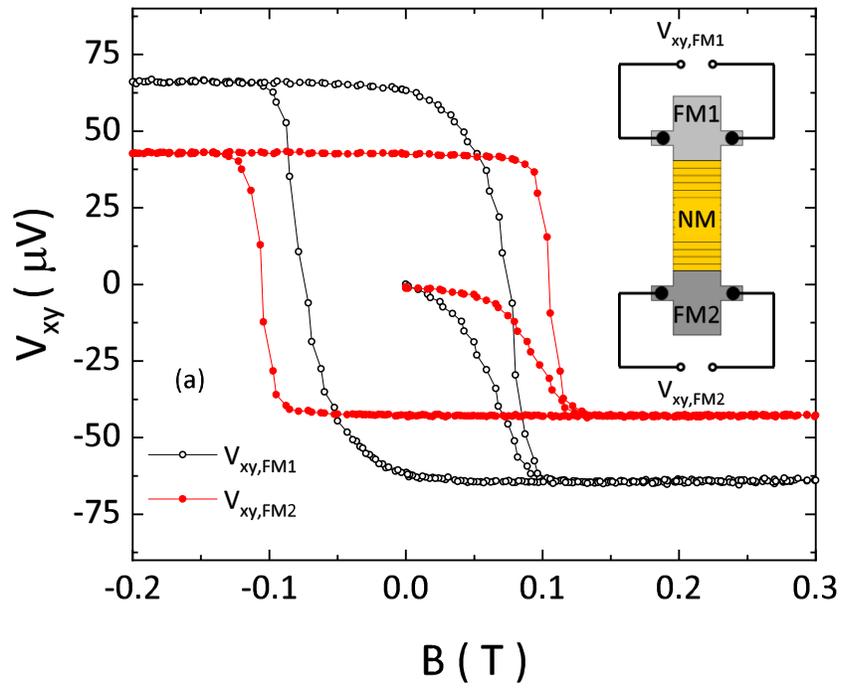

Fig. 5a

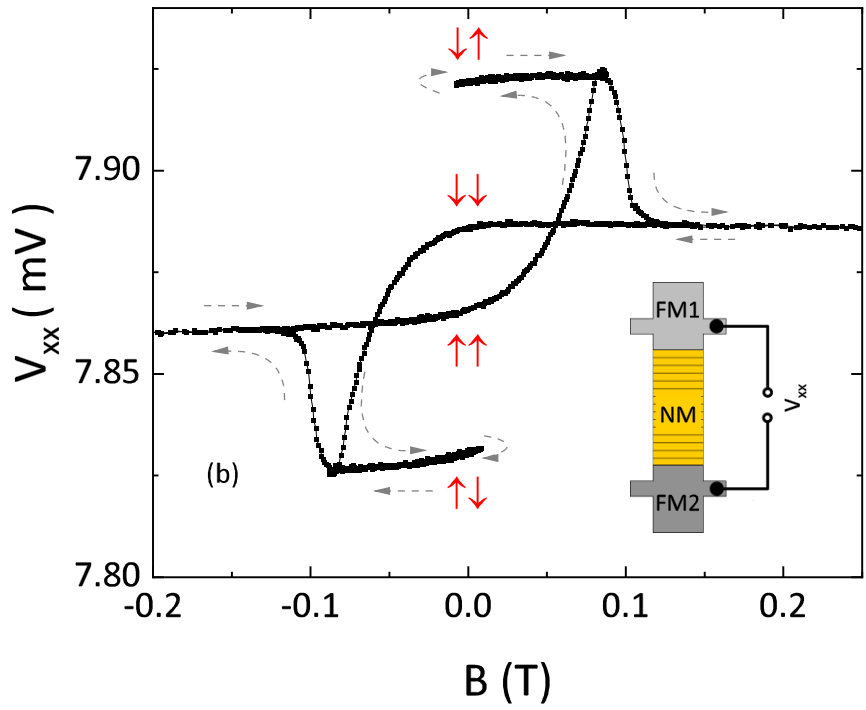

Fig.5b